# Optimizing Portfolio Management and Risk Assessment in Digital Assets Using Deep Learning for Predictive Analysis

Qishuo Cheng[1,*]

Department of Economics
University of Chicago
Chicago, IL, USA
qishuoc@uchicago.edu

Le Yang[2]

Computer Information Science
Sam Houston State University
Huntsville，TX，USA
wesleyyang96@gmail.com

Jiajian Zheng[3]

Bachelor of Engineering
Guangdong University of Technology
ShenZhen,CN
im.jiajianzheng@gmail.com

Miao Tian[4]

Master of Science in Computer Science
San Fransisco Bay University
Fremont CA, USA
miao.hnlk@gmail.com

Duan Xin[5]

Accounting
Sun Yat-Sen University
HongKong
duanxin12314057@gmail.com

**Abstract.** Portfolio management issues have been extensively studied in the field of artificial intelligence in recent years, but existing deep learning-based quantitative trading methods have some areas where they could be improved. First of all, the prediction mode of stocks is singular; often, only one trading expert is trained by a model, and the trading decision is solely based on the prediction results of the model. Secondly, the data source used by the model is relatively simple, and only considers the data of the stock itself, ignoring the impact of the whole market risk on the stock. In this paper, the DQN algorithm is introduced into asset management portfolios in a novel and straightforward way, and the performance greatly exceeds the benchmark, which fully proves the effectiveness of the DRL algorithm in portfolio management. This also inspires us to consider the complexity of financial problems, and the use of algorithms should be fully combined with the problems to adapt. Finally, in this paper, the strategy is implemented by selecting the assets and actions with the largest Q value. Since different assets are trained separately as environments, there may be a phenomenon of Q value drift among different assets (different assets have different Q value distribution areas), which may easily lead to incorrect asset selection. Consider adding constraints so that the Q values of different assets share a Q value distribution to improve results.

---

1 * Corresponding author: [Qishuo Cheng]. Email: [qishuoc@uchicago.edu].

# 1   Introduction

With the development of artificial intelligence, much progress has been made in the study of portfolio management problems, aiming to maximize the expected return of multiple risky assets. Simultaneously, China's stock exchange market is gradually evolving towards diversification, convenience, and information, resulting in the generation of vast amounts of data daily. To address the deficiency in traditional transaction analysis methods, which struggle with handling large datasets, and to mitigate the irrational operations of undisciplined human investors, quantitative investment characterized by scientific, systematic, and accurate approaches has gradually become a focal point for institutional investment researchers.

Additionally, the trends of certain individual stocks exhibit a correlation with the overall market trend. This study expands the application of deep reinforcement learning in the field of quantitative investment, offering significant references and practical guidance for the integration of deep reinforcement learning in financial investment, particularly in stock investment.

# 2   Related Work

## 2.1   Portfolio management

The portfolio management department, a key component of risk control, handles risk measurement and data analysis, reporting to top executives. This article explores portfolio management's integral role in the product or customer life cycle, emphasizing risk management through indicators and discussing monitoring, forecasting, and early warning systems. This cohesive process underscores the department's significance in asset portfolio management. The construction of monitoring forecast and early warning will mainly focus on the prediction methods of common indicators, and how we monitor forecast and early warning, these three blocks are linked together, mainly the construction and combing of a process, asset portfolio management, asset portfolio management.

## 2.2 Deep learning and Deep Q-learning

The application of deep reinforcement learning in quantitative investment not only expands research ideas and modeling methods but also drives financial transactions towards intelligence and automation. It facilitates the identification of optimal investment opportunities and trading signals, adapts to diverse stock trading environments based on varying market conditions, and enhances the prediction accuracy of different assets.

### 2.2.1 Deep Q-learning.

Deep Q-Learning algorithm is referred to as DQN. DQN evolved on the basis of Q-Learning. DQN's modification of Q-Learning mainly includes two aspects:
1) DQN uses deep convolutional neural networks to approximate value functions
2) DQN uses experiential replay of the learning process of training reinforcement learning

### 2.2.2 The structure is as follows:

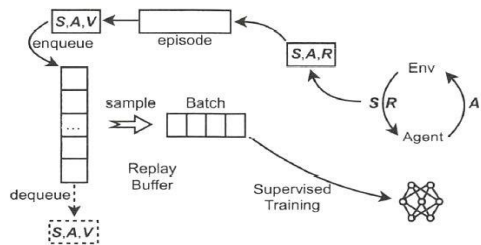

**Fig. 1.** Deep Q-Network

(1)The DQN approximates the behavior value function using a neural network, constituting a nonlinear approximation. The network structure comprises three convolutional layers and two fully connected layers. In formulaic expression, the value function is denoted as $Q(s,a; \omega)$, where updating the network effectively means updating the parameters, and once $\omega$ is set, the network parameters are fixed.

(2)A key feature of DQN is the introduction of experience replay. In contrast to the traditional Q-Learning method, which iterates and improves based on the current strategy, DQN employs data generated through interaction for learning. Unlike tabular models, using machine learning models in online learning may encounter two related problems. The sequence derived from interaction exhibits some correlation, violating the crucial assumption of independence and identical distribution in maximum likelihood-based machine learning models. This correlation disrupts the independence assumption, leading to significant fluctuations in the learned value function model.

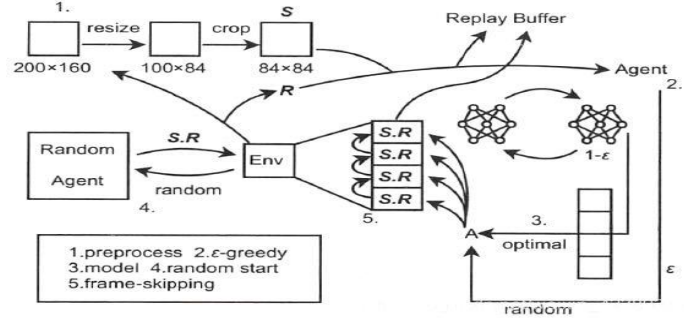

**Fig. 2**. Replay Buffer quintuple model

Efficiency of interactive data use. When the gradient descent method is used to update the model, the model training often needs to converge after several iterations. Each iteration requires a certain number of samples to calculate the gradient, and if each calculated sample is discarded after one gradient calculation, then we need to spend more time interacting with the environment and collecting samples.

## 3 Methodology

In this study, each asset is treated as an environment, and the strategy is trained by considering the income from holding the asset as its return and the average income from other portfolio assets as the cash return. The goal is to allocate cash to assets with expected performance above the average. The method is applied to 48 US stock portfolios, varying from 10 to 500 stocks, with diverse selection criteria and transaction costs. The algorithm employs a single hyperparameter setting across all portfolios, demonstrating superior performance compared to passive and active benchmark investment strategies.

### 3.1 Algorithm establishment

The algorithm in this paper mainly needs to be explained:
1. Train each asset separately as an environment, sampling one asset initialization environment at a time;
2. Use the Q function of cumulative return evaluation training on the verification data, and record the Q function with the best performance on the verification data;
3. The action space is a two-dimensional discrete action, with 0 representing holding cash and 1 representing holding the asset.
The main algorithm formula is as follows:

$$\nabla\theta_i L_i(\theta_i) = E_{s'\sim\varepsilon}[R_t + \gamma_{a'}^{max} Q(s',a';\theta_{i-1}) - Q(s,a;\theta_i)\nabla\theta_i Q(s,a;\theta_i)] \quad (1)$$

$$R_t = \{r_{t+1,i} - (1 - a_{t-1})C \ if \ a_t = 1 \quad (2)$$

### 3.2 Data result

Parameter setting：

**Table 1.** List of hyperparameters used for all portfolio setups,

| Parameter | Symbol | Value |
|---|---|---|
| Discount factor | $\gamma$ | 0.9 |
| Exploration probability | $\epsilon$ | 0.3 |
| Number of iterations | N | 3,000,000 |
| Experience memory size | N | 300,000 |
| Gradient step interval | - | 20 |
| Evaluation interval | $\Omega$ | 10,000 |
| Optimizer | - | Adam |
| Batch size | - | 1024 |
| Number of hidden layers | - | 2 |
| Number of neuronsin hidden layers | - | 32or64or128 |

### 3.3 Test data

The algorithm selected the historical data of 500 US stocks from 2010-01-01 to 2021-06-30 for testing.
Data division:
Training data set period: 2010-01-01 to 2018-12-31
Data set validation period: 2019-01-01 to 2019-31
Test data set period: 2020-01-01 to 2021-06-30
Portfolio grouping: Test 500 stocks by market capitalization as low, mid, and high.
Transaction cost: bps
Test method: Three agents were trained on the training set for model integration and tested on the test set.
Comparison basis:
Buy-and-hold: equal rights to Buy and hold all shares;
Momentum: A strategy of buying stocks that have had a positive average return over the past five trading days;
Reversion: A return strategy to buy stocks that have had negative average returns over the past five trading days.

6**Table 2.** Out-of-sample cumulative returns at the end of the test period for each level of transaction costs and different portfolios

| Transaction costs | Portfolio size | Portfolio type | Agent | Buy-and-hold | Momentum | Reversion |
|---|---|---|---|---|---|---|
| 1 bp | 10 | big | 222.9% | 129.5% | 103.6% | 63.8% |
| | | random | 68.1% | 56.4% | 50.6% | 29.7% |
| | | small | 109.0% | 65.7% | 100.7% | 108.7% |
| | 25 | big | 41.4% | 66.2% | 39.4% | 58.8% |
| | | random | 78.6% | 48.1% | 47.6% | 38.2% |
| | | small | 147.8% | 84.0% | 108.5% | 94.1% |
| | 50 | big | 101.0% | 52.7% | 44.5% | 60.9% |
| | | random | 90.8% | 62.5% | 16.9% | 79.9% |
| | | small | 119.0% | 66.9% | 51.7% | 85.9% |
| | 100 | big | 126.0% | 56.3% | 53.8% | 90.1% |
| | | random | 185.9% | 67.8% | 17.1% | 116.6% |
| | | small | 475.0% | 68.0% | 52.2% | 72.0% |
| | 200 | big | 115.6% | 58.9% | 31.0% | 111.5% |
| | | random | 84.4% | 63.6% | 17.7% | 106.1% |
| | | small | 195.0% | 59.4% | 41.0% | 76.5% |
| | 500 | all | 155.1% | 59.2% | 40.4% | 89.5% |
| | Mean | | 144.7% | 66.6% | 51.1% | 80.1% |
| 5 bps | 10 | big | 123.2% | 129.5% | 76.6% | 44.3% |
| | | random | 24.2% | 56.4% | 31.1% | 13.1% |
| | | small | 173.7% | 65.7% | 73.6% | 81.7% |
| | 25 | big | 11.6% | 66.2% | 20.4% | 37.1% |
| | | random | 38.5% | 48.1% | 27.4% | 19.2% |
| | | small | 134.6% | 84.0% | 79.9% | 67.5% |
| | 50 | big | 76.0% | 52.7% | 24.7% | 38.8% |
| | | random | 79.2% | 62.5% | 0.6% | 54.8% |
| | | small | 64.3% | 66.9% | 30.7% | 60.2% |
| | 100 | big | 104.4% | 56.3% | 32.5% | 63.6% |
| | | random | 107.3% | 67.8% | 0.8% | 86.5% |
| | | small | 204.1% | 68.0% | 30.9% | 48.0% |
| | 200 | big | 89.0% | 58.9% | 12.8% | 82.0% |
| | | random | 103.1% | 63.6% | 1.3% | 77.4% |
| | | small | 99.0% | 59.4% | 21.3% | 51.8% |
| | 500 | all | 100.8% | 59.2% | 20.8% | 63.1% |
| | Mean | | 95.8% | 66.6% | 30.3% | 55.6% |
| 10 bps | 10 | big | 127.6% | 129.5% | 47.8% | 23.3% |
| | | random | 3.5% | 56.4% | 10.2% | -4.7% |
| | | small | 68.3% | 65.7% | 44.7% | 52.7% |
| | 25 | big | 46.1% | 66.2% | 0.2% | 14.2% |
| | | random | -41.0% | 48.1% | 5.8% | -0.9% |
| | | small | 114.3% | 84.0% | 49.5% | 39.3% |
| | 50 | big | 53.8% | 52.7% | 3.6% | 15.4% |
| | | random | 55.2% | 62.5% | -16.6% | 28.4% |
| | | small | 107.7%6 | 66.9% | 8.5% | 32.9% |
| | 100 | big | 69.4% | 56.3% | 9.9% | 35.7% |
|  |  | random | 131.0% | 67.8% | -16.4% | 54.7% |
|  |  | small | 153.1% | 68.0% | 8.5% | 22.6% |
|  | 200 | big | 74.1% | 58.9% | -6.5% | 50.8% |
|  |  | random | 70.2%6 | 63.6% | -16.0% | 47.0% |
|  |  | small | 92.7% | 59.4% | 0.5% | 25.8% |
|  | 500 | all | 86.0% | 59.2% | 0.1% | 35.1% |
|  | Mean |  | 75.7% | 66.6% | 8.4% | 29.5% |

Analysis:
Among 48 groups of experiments, DQN was the best in 36 groups of experiments.
The larger the portfolio, the more stocks to choose from, the better the performance of DQN strategy;
In a small-cap portfolio, the DQN strategy outperforms the benchmark strategy.
Compared with momentum and regression strategies, DQN strategies have better applicability to transaction costs.

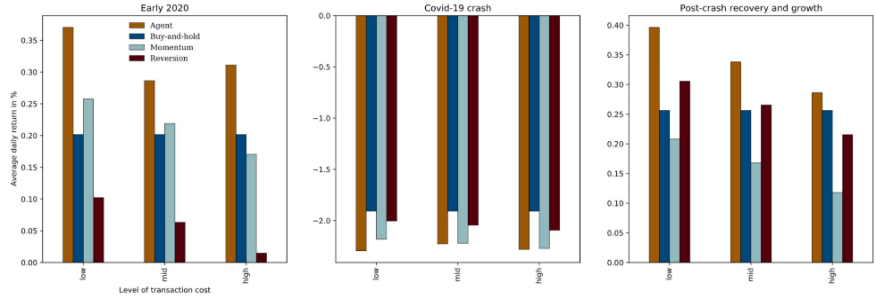

**Fig. 3.** Out-of-sample trading performa three phases based on market conditions

The whole test stage can be divided into three parts: the normal market stage before the COVID-19 epidemic, the market decline stage when COVID-19 appears, and the market recovery stage after. The DQN strategy performed optimally in both the normal phase and the recovery phase, but in the market decline phase, DQN performed poorly, possibly because there was no similar situation in the training data.

### 3.4 Experimental conclusion

In this study, the application of DQN to asset management portfolios involves randomly selecting a single asset from a group for trading as an environment. The reward calculation, based on the average return of the asset group, establishes a connection between the transaction return in a single asset environment and the portfolio return. With a single hyperparameter setting, the trained strategy demonstrates adaptability to various trading scenarios, achieving optimal performance in 36 out of 48 experimental groups. However, the strategy experiences significant performance degradation in emergency situations when compared to benchmark methods.
Nevertheless, the experiment reveals certain limitations. Firstly, while introducing the DQN algorithm into asset management proves effective, inspiring confidence in the efficacy of DRL in portfolio management, the study emphasizes the need for adapta-tion to the complexity of financial problems. Secondly, the strategy relies on selecting assets and actions with the highest Q value, leading to potential Q value drift among different assets trained separately as environments. To mitigate this issue, the suggestion is to consider incorporating constraints to ensure a shared Q value distribution among different assets for improved results.

## 4   Conclusion

In this study, we successfully introduce deep reinforcement learning (DRL) into the field of asset portfolio management, and use deep Q learning (DQN) algorithm to make asset trading decisions. The experimental results show that DQN strategy performs best in 36 out of 48 experiments, which provides an effective method for asset portfolio management. However, the experiment also revealed some shortcomings. In an emergency, the strategy could face performance degradation, suggesting that we need to think more about algorithmic robustness for complex financial problems. For future research directions, we believe it is necessary to improve the limitations of current strategies. First, we should consider introducing constraints in the asset selection process to prevent Q-value drift between different assets, thereby improving the robustness and accuracy of the strategy. Second, we need to delve into the reasons why strategy performance deteriorates in emergency market situations and look for ways to improve. Finally, we encourage the application of deep reinforcement learning to broader areas of finance, such as risk management and forecasting, to explore more potential applications in finance.

Overall, the application of deep reinforcement learning in portfolio management and risk management shows great potential, providing new ideas and methods for innovation in the financial sector. By continuously improving algorithms and expanding application areas, we are expected to better leverage the benefits of deep reinforcement learning to achieve smarter and more accurate asset management and risk assessment.

## 5   ACKNOWLEDGEMENT

During the completion of this study, I would like to sincerely thank Liu, Bo and others for their outstanding contributions to the academic community. Their journal article "Integration and Performance Analysis of Artificial Intelligence and Computer Vision Based on Deep Learning. Algorithms provides valuable inspiration and enlightenment for my research on financial portfolio management and risk prediction in the field of machine reinforcement learning. This paper deeply discusses the integration and performance analysis of artificial intelligence and computer vision based on deep learning algorithms, providing me with rich theoretical support in experimental design and reinforcement learning application. Some of the core conclusions and methods adopted in my research are derived from this article, which has played an important guiding role for my research.